\documentclass[runningheads]{llncs}
\usepackage{mathtools}
\usepackage{cite}
\usepackage{caption}
\usepackage{amsmath,amssymb,amsfonts}
\usepackage{algorithmic}
\usepackage{graphicx}
\usepackage{textcomp}
\usepackage{xcolor}
\usepackage{booktabs}
\usepackage{multirow}
\usepackage{ulem}
\usepackage{url}
\usepackage{hyperref}
\usepackage{amsmath}
\usepackage{stfloats}
\usepackage[misc]{ifsym}
\usepackage{MnSymbol}
\usepackage{cuted}
\usepackage{multicol}
\usepackage{lipsum}

\newtheorem{assumption}{Assumption}
\hypersetup{
colorlinks=true,
linkcolor=black
}
\begin{document}
\title{Separating and Learning Latent Confounders to Enhancing User Preferences Modeling}
\titlerunning{Separating and Learning Latent Confounders.}
% \author{Anonymous author(s)}
% \institute{}
\author{
Hangtong Xu \and
Yuanbo Xu \Letter \and
Yongjian Yang
}
\authorrunning{Y. Xu et al.}
\institute{MIC Lab, College of Computer Science and Technology, Jilin University
\email{\{yuanbox,yyj\}@jlu.edu.cn}, \email{xuht21@mails.jlu.edu.cn}
}
\maketitle              % typeset the header of the contribution
\begin{abstract}
Recommender models aim to capture user preferences from historical feedback and then predict user-specific feedback on candidate items. However, the presence of various unmeasured confounders causes deviations between the user preferences in the historical feedback and the true preferences, resulting in models not meeting their expected performance. Existing debias models either (1) specific to solving one particular bias or (2) directly obtain auxiliary information from user historical feedback, which cannot identify whether the learned preferences are true user preferences or mixed with unmeasured confounders. Moreover, we find that the former recommender system is not only a successor to unmeasured confounders but also acts as an unmeasured confounder affecting user preference modeling, which has always been neglected in previous studies. To this end, we incorporate the effect of the former recommender system and treat it as a proxy for all unmeasured confounders. We propose a novel framework, \textbf{S}eparating and \textbf{L}earning Latent Confounders \textbf{F}or \textbf{R}ecommendation (\textbf{SLFR}), which obtains the representation of unmeasured confounders to identify the counterfactual feedback by disentangling user preferences and unmeasured confounders, then guides the target model to capture the true preferences of users. Extensive experiments in five real-world datasets validate the advantages of our method.
\keywords{Recommender system  \and Debiasing  \and Causal inference  \and Preference Modeling}
\end{abstract}
\section{Introduction}
\label{section:1}
Recommender systems play an essential role in mitigating information explosion and a wide range of real-world applications. Such systems basically aim to match a user with her favorite items, and a large number of recommender models have been proposed in the research community to achieve this goal \cite{xu1,xu2}. However, there are various types of biases that occur during the generation of user feedback data, resulting in the sub-optimal performance of the model. For example, people only access a small number of items in real-world scenarios (i.e. exposure bias\cite{exposure}), and popular items may have higher chances to be clicked (i.e. popularity bias\cite{popularity}); incomplete user interests drive the personalized system to generate some homogeneous candidates (i.e. homogeneous\cite{homogeneous,xu3,xu4}). Existing recommender models tend to ignore the bias in the feedback data during the design stage and simply match the data during training, thus performing poorly once the model is used in real-world scenarios.\par
In recent years, debiasing has become one of the research foci in the recommender system field. Various debiasing methods have been proposed, and among them, specific for one particular bias is widely used. However, various biases present a mixture in the data, thus debiasing methods for a specific bias can improve the performance of the model but still fall short of our expectations. With the rise of causal learning in recommender systems, the use of causal analysis to analyze the causes of bias and then employ causal tools for debiasing has been widely proposed. Nevertheless, such methods require auxiliary information about the confounders from the dataset before they are used, which is against the fact that most of the confounders in real scenarios are unmeasured, and thus have enormous limitations. Moreover, most existing debiasing methods overlook the influence of the recommender system on the user preferences and only take the confounders as the cause of the bias. Feedback data collection is mainly from records of user interactions on items recommended by the recommender system, in which the recommender system plays an essential role, it is not advisable to ignore the impact of recommender systems. It is rare to accurately recommend a list of items that perfectly match the preferences of the user, typically a suboptimal list of items. The feedback data generated by the interaction between the user and the suboptimal item list can not reflect the true preferences of the user, and recommender systems trained on these data further exacerbate the preference bias. A more extreme scenario is that when recommendation strategies of a recommender system increase exposure to a certain kind of item, the number of positive feedbacks for that item in the feedback data will be significantly higher (e.g., higher sales of the item due to promotional strategies). Thus the new recommender system will conclude that more users like this kind of item, making the user preferences more deviated from the true preferences.\par
To mitigate the influence of unmeasured confounders and former recommender systems on user preferences, we use causal analysis to thoroughly analyze the causes of user preference deviation. We further find that the recommender system inherits the influence from confounders and continuously exacerbates the user preference deviation with the recommender system update iterations. As the examples given earlier, the causes of the influence of recommender systems on user preferences are usually user-independent (e.g., recommendation policies), and most of the unmeasured confounders are independent of the user, which provides the necessary premise for further separation of user preferences and confounders. Based on the connection between the confounders and the former recommender system, we use the former recommender system as their proxy and use variational inference to separate the confounders and user preferences in the latent parameter space to obtain the representation of the confounders. Then we propose a novel framework, named \textbf{SLFR}, which obtains the representation of latent confounders to identify the counterfactual feedback by disentangling user preferences and latent confounders, then guides the model to capture the true preferences of users. The proposed SLFR is model-agnostic, which can be easily deployed in any state-of-the-art recommendation model. Extensive experimental results from five real-world datasets demonstrate that our method outperforms baselines and can be easily used in other state-of-the-art models for performance improvement. The contributions of this paper are summarized as follows:\par
\textbf{(1)} We investigate the new problem of debiasing in recommender systems when incorporating the effects of former recommender systems and unmeasured confounders. \textbf{(2)} We state the assumption of independence of confounders and user preferences, the basis for separating them in the latent parameter space. \textbf{(3)} We propose a novel framework, Separating and Learning Latent Confounders For Recommendation (SLFR), which obtains the representation of latent confounders to assist the model in capturing the true preferences of users. \textbf{(4)} We conduct extensive experiments that include both general and specific debiasing scenarios to validate the advantages of our method.\par
\begin{figure}[!t]
\centering
\caption{Causal Graph Views of Recommendation Problem. (a): General view without unmeasured confounders; (b): The view that takes into account unmeasured confounders and former recommender system; (c): The view uses the former recommender system as the proxy for confounders.}
\label{fig:2}
\includegraphics[width = 0.7\linewidth]{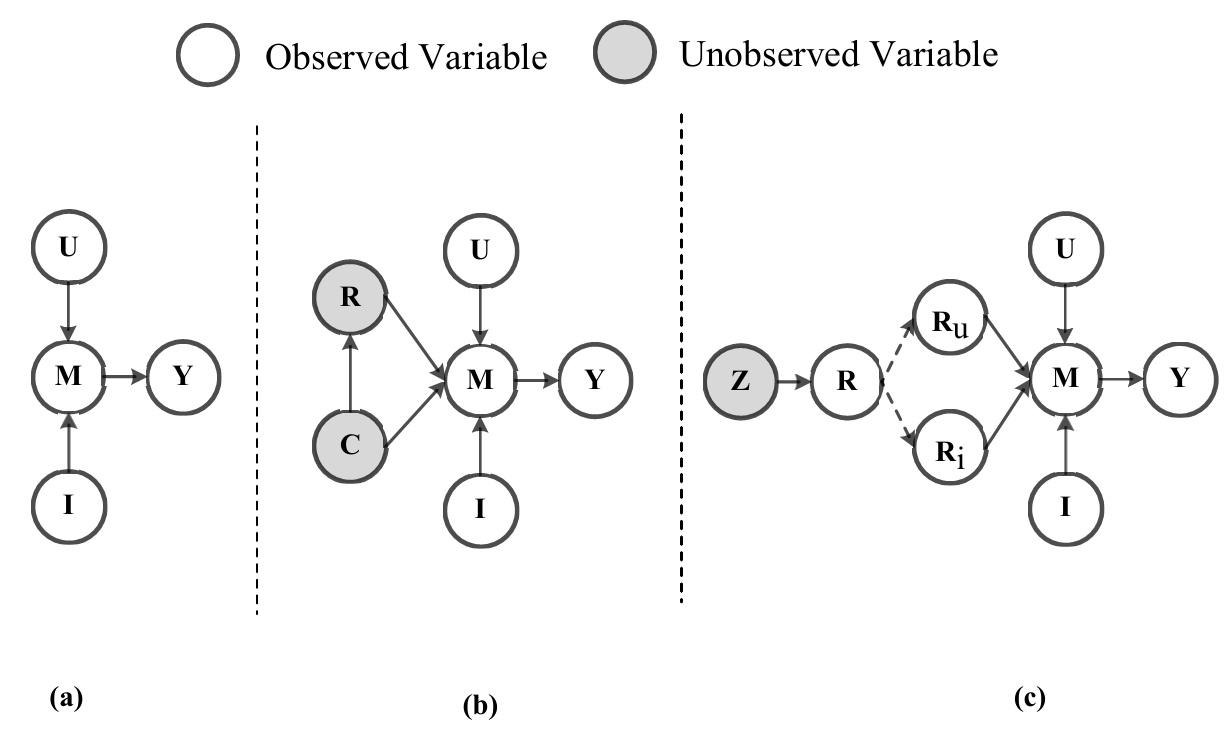}
\end{figure}
\section{Problem Formulation}
% In this section, we first analyze the recommendation problem from a causal point of view in two cases, the presence or absence of unmeasured confounders. Then we analyze a real-world dataset to verify the impact of the former recommender system on user feedback. Finally, we redefine the recommendation problem by introducing the former recommender system as a proxy for unmeasured confounders\par
\subsection{Usual View of Recommendation Problem}
To study the recommendation problem, we define a causal graph to describe the generation of user feedback in Figure \ref{fig:2}. Note that we use capital letters (e.g., \textbf{U}) to denote a variable, lowercase letters (e.g., $\mathit{u}$) to denote its specific value, and calligraphic font letters (e.g., $\mathcal{U}$) denote sample space. In particular, \textbf{U} indicates the user representation, \textbf{I} indicates the item representation, \textbf{Y} indicates the user feedback on the item in the feedback data, \textbf{C} indicates the set of latent confounders that may cause user preference deviation in the feedback data, \textbf{R} indicate the special confounder: the former recommender system, which has a direct impact on user-item interactions (e.g., recommendation strategies). \textbf{Z} indicate the latent representation space of unmeasured confounders, which can be divided into two categories, $\textbf{R}_u$: impact on items (e.g., exposure). $\textbf{R}_i$: impact on users (e.g., homogeneous). \textbf{M} indicates the matching interaction process between users and items.\\\par
\noindent\textbf{General Recommendation Problem.} Given observation data, a recommendation algorithm aims to accurately predict the feedback of user $\mathit{u}$ on item $\mathit{i}$, if the item is in the set of interaction items of $\mathit{u}$ in the dataset.
\begin{equation}
    P(\hat{Y} = 1 \mid \mathit{u ,i}) = p(Y = 1 \mid \mathit{u ,i}).
\end{equation}
After the model reaches the optimum, for a particular user, the model ranks the items based on the $P(\hat{Y} = 1 \mid \mathit{u, i})$ and outputs the final recommendation list. The problem is defined based on the assumption that the user-item interaction process depends only on the user and the item itself, and is not influenced by other factors, the causal graph is shown in Figure \ref{fig:2} (a). However, in real-world scenarios, users interact with the recommended items without randomized controlled trials, inevitably affected by unmeasured confounders, thus there is a deviation between the user preferences inherent in the feedback data and the true user preferences, the causal graph is shown in Figure \ref{fig:2} (b).\par
\begin{figure}[!t]
\centering
\caption{Analysis of the Reasoner dataset. After converting feedback to implicit feedback, we summarize as follows, left: false positive and false negative percent in Reasoner; right: percentage difference between positive and negative feedback for the Kuairec and Reasoner datasets.}
\label{fig:3}
\includegraphics[width = 0.8\linewidth]{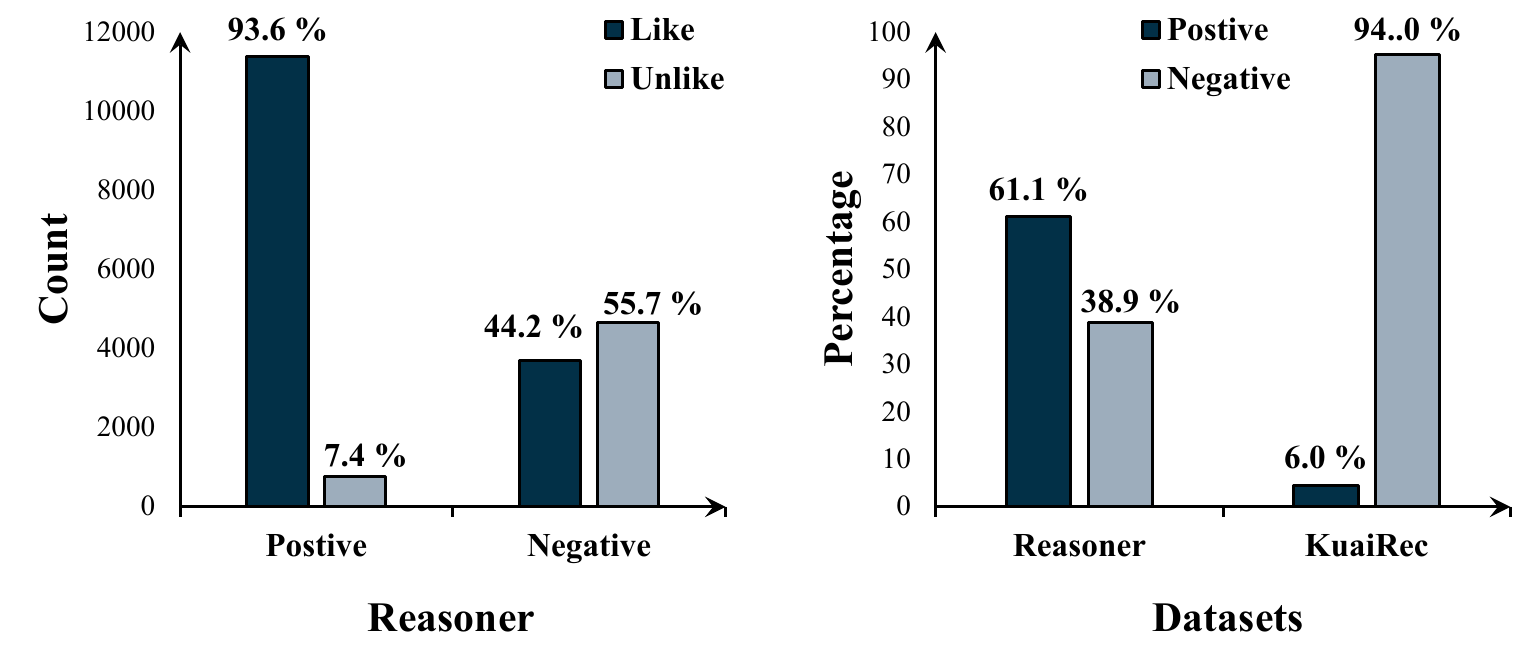}
\end{figure}
\subsection{General Debiasing Recommendation Problem}
Previous work has used \textbf{C} as a specific factor, such as popularity, video time, and the specific feature of items, which can usually be obtained from feedback data but may result in a large variance in model performance due to the inability to calculate accurately. Moreover, there are more unmeasured confounders besides the ones mentioned above, \textbf{C} is a latent variable represented by a shaded node in Figure \ref{fig:2} (b). Further, we review the feedback data generation process and find that the former recommender system plays an important role, as we discussed in Section \ref{section:1} for specific examples. To validate our idea, we perform a detailed analysis on a real-world dataset.\par
Reasonear \cite{reasoner} is a real-world dataset containing both user ratings and true preference label data, which can help us to verify whether the feedback collected from the former recommender system matches the true preferences of users for the same items. To make the analysis more general, we take the ratings $\geq 4$ as positive feedback, and others as negative feedback, and Kuairec \cite{kuairec} is used as a representative of the general dataset for comparison. It is important to emphasize that the items in the dataset are fully exposed, even those with negative feedback. The results are shown in Figure \ref{fig:3}. It is clear from Figure \ref{fig:3} (left) that a large proportion of the items in the negative feedback have a true preference of the user for Like, even in the positive feedback there is a proportion of users whose true preference is the opposite of the label in the dataset. Figure \ref{fig:3} (right) that the percentage of positive feedback is much higher in the Reasoner dataset than in the usual dataset, implying that the feedback-true preference false matching problem due to the presence of the former recommender system is even more severe in the usual datasets.\par
In addition, previous work \cite{bias_amp,feedbackloop,xu5} has found that biases accumulate and amplify during recommender system iterations, and we believe that this may be due to the fact that the recommender system not only inherits the influence of the existing confounders but also acts as a special confounder to intensify the deviation of user preferences. We further split it into two categories of influence users ($R_{U}$) and items ($R_{I}$) in Figure \ref{fig:2} (c), respectively. Thus, our work focuses on the problem setting where the unmeasured confounders are present and incorporates the former recommender system as a special unmeasured confounder, which is a more general problem.\par
\begin{problem}[General Debiasing Recommendation Problem]
\label{problem}
Given observation data collection from the former recommender system, the task is to capture the pure user preference from heterogeneous preference inside data.
\end{problem}
\section{Method}
% In this section, we first give a causal view of the causes of bias in user preferences, and then propose a novel approach to separate and learn latent confounders from the latent parameter space.  We present the framework of STPR based on the confounder representation.
\subsection{Causal View of Bias in Recommendation}
As we discussed above, the former recommender system succeeds the effects of the confounders and deepens the deviation of user preferences as a new confounder, thus sharing a unique latent parameter space with the other confounders. We, therefore, chose to take the former recommender system as a proxy for all confounders ($\textbf{C} \subseteq \textbf{R}$), and the causal graph is shown in Figure \ref{fig:2} (c). As Figure \ref{fig:2} (c) shows, due to the confounders(\textbf{R}), existing recommender system models that estimate the conditional probability $P(Y \mid U, I)$ inevitably affected by confounders, which leads to deviation of user preference. Formally, given U = $\mathit{u}$ and I = $\mathit{i}$, we can derive the conditional probability $P(Y \mid U,I)$ by:
\begin{equation}
    \label{eq:2}
    \begin{aligned}
     P(Y \vert U = u,I = i) &= \sum_{\mathit{r} \in \mathcal{R}}\sum_{\mathit{m} \in \mathcal{M}} \frac{P(u)P(i)P(r\vert z)P(m \vert u,i,r)P(Y \vert m)}{P(u)P(i)}  \\
     &= P(r_{u})P(r_{i})P(Y \vert M(u,i,r_{u},r_{i}))),
\end{aligned}
\end{equation}
where $\mathcal{R}$ and $\mathcal{M}$ are the sample space of \textbf{R} and \textbf{M},respectively. $\mathit{z}$ is the latent parameter space of confounders, we further split it into two categories of influence users ($\mathit{R}_{U}$) and items ($\mathit{R}_{I}$) to further solve the problem.\par
From Eq. \ref{eq:2}, we can find that \textbf{R} does not only affect the final observe feedback data Y but also the user and item preference modeling via $M(u, i,r_{z})$. For example, popular items under particular confounders will have higher exposure, thus the expectation of feedback score $E[Y]$ will be higher than unpopularity ones, in turn, the user exposed to popular items may give the model an incorrect signal (the popular items are favored by users), higher $E[P(Y = 1 \mid U = u, I = i)]$ for popular items. However, since most of the confounders are not measurable, making it impossible to obtain the information directly from the dataset. A possible solution is to learn a representation of the unmeasured confounders from the latent variable space and, based on this representation, obtain a counterfactual representation of the true preferences of the user.\par

\subsection{Separating and Learning Latent Confounders}
There are a wide variety of unmeasured confounders, and we note that most of them are not user-dependent or item-dependent, such as exposure strategy, which is correlated with item exposures but relatively independent of the user. We further separate the effects of confounders into $R_u$ and $R_i$ based on the above relatively independent relations to encompass as much of the confounders as possible. Before we present a further solution to the problem, we first give the following assumption:
\begin{assumption}[Independence of unmeasured confounders]
Given the unmeasured confounder $r_u\in R_U, r_i \in R_I $, user $u$ and item $i$, we have $r_u \upmodels u$ and $r_i \upmodels i$.
\end{assumption}
Based on the assumption, we are able to separate the parameter latent space into two parts, one for confounders and one for users (items). The problem is transformed into how to learn the representation of the confounders from the parameter latent space. Inspired by VAE, which is widely used to learn latent variable representations of data, we try to use variational inference to learn representations of unmeasured confounders. The normal VAE loss function can be written as follows:
\begin{equation}
    \mathcal{L}_{VAE} = \textbf{E}_{z \sim q_{\phi}(z \vert x)} \log p_{\theta}(x \vert z) - \textbf{KL}(q_{\phi}(z \vert x) \Vert p_{\theta}(z)).
\end{equation}
The VAE can learn a representation of the latent parameter space from input data. However, for a recommender system, the learned representation is a mixture of user preferences (item preferences) and confounders, not the pure representation of user preferences we want, hence direct use of the VAE does not perform very well. After further unpacking the loss function of VAE, we find some surprising properties:
\begin{equation}
    \begin{small}
    \begin{aligned}
        &\textbf{KL}(q_{\phi}(z \vert x) \Vert p_{\theta}(z)) = \frac{1}{N} \sum\limits^{N}\limits_{u = 1}\textbf{KL}(q(z \vert x_{u}) \Vert p(z)) \\
        &= \textbf{KL}(q(z,u) \Vert q(z)p(u)) + \textbf{KL}(q(z) \Vert \prod_{j}q(z_{j})) + \textbf{KL}(q(z_{j}) \Vert p(z_{j})).
    \end{aligned}
    \end{small}
\end{equation}
It includes three parts(from left to right): The first part represents the latent variables and user associations, increasing the weight of the first part minimizes the correlation between the user and the latent variable, ensures the independence of the latent variable, and thus allows for latent variables that are not relevant to users; The second part is for reducing the correlation between the dimensions of the latent variables and thus increasing the amount of information that the latent variables can contain; The last part is to guarantee the relevance of the latent variables to the data, ensuring that the learned latent variable representations are drawn from the latent parameter space of the data and are not randomly generated. Thus, in order to learn the desired user-independent (item-independent) confounders representation, we make the following modifications to the loss function:
\begin{equation}
    \label{eq:5}
    \begin{small}
    \begin{aligned}
        &\textbf{KL}_\alpha(q_{\phi}(z \vert x) \Vert p_{\theta}(z))\\
        &=\alpha\textbf{KL}(q(z,u) \Vert q(z)p(u)) + \textbf{KL}(q(z) \Vert \prod_{j}q(z_{j})) + \sum_{j} \textbf{KL}(q(z_{j}) \Vert p(z_{j})),\\
        &\mathcal{L}_{VAEBlock} = \textbf{E}_{z \sim q_{\phi}(z \vert x)} \log p_{\theta}(x \vert z) - \textbf{KL}_\alpha(q_{\phi}(z \vert x) \Vert p_{\theta}(z)).
    \end{aligned}
    \end{small}
\end{equation}
By applying the loss function shown in Eq.\ref{eq:5}, we can have representations $q_{\phi_u}(r_u \mid x)$, $q_{\phi_i}(r_i \mid x)$ of the unmeasured confounders that are independent of user and item.
\subsection{Framework}
In order to address the General Debiasing Recommendation Problem, we propose a novel debiasing framework named SLFR, which consists of two stages.\\\par
\noindent\textbf{Learning representation of unmeasured confounders.} In this stage, we pre-train two VAE models based on the loss function Eq.\ref{eq:5} to learn the unmeasured confounders representing $r_u$ and $r_i$ for users and items, respectively, to indicate the different impact of confounders on users and items. It is necessary to emphasize that the optimal state of the model is to learn the representation of the confounders completely independently of the user, but this is difficult to achieve. Therefore, we learn the representation of the confounders separately for each user to ensure that the final result is as expected. For $ U = u, I = i $, we have the representation of confounders $r_u$ and $r_i$, respectively.\\\par
\begin{figure}[!t]
\centering
\caption{The Framework of SLFR under the backbone MF.}
\label{fig:4}
\includegraphics[width = 0.8\linewidth]{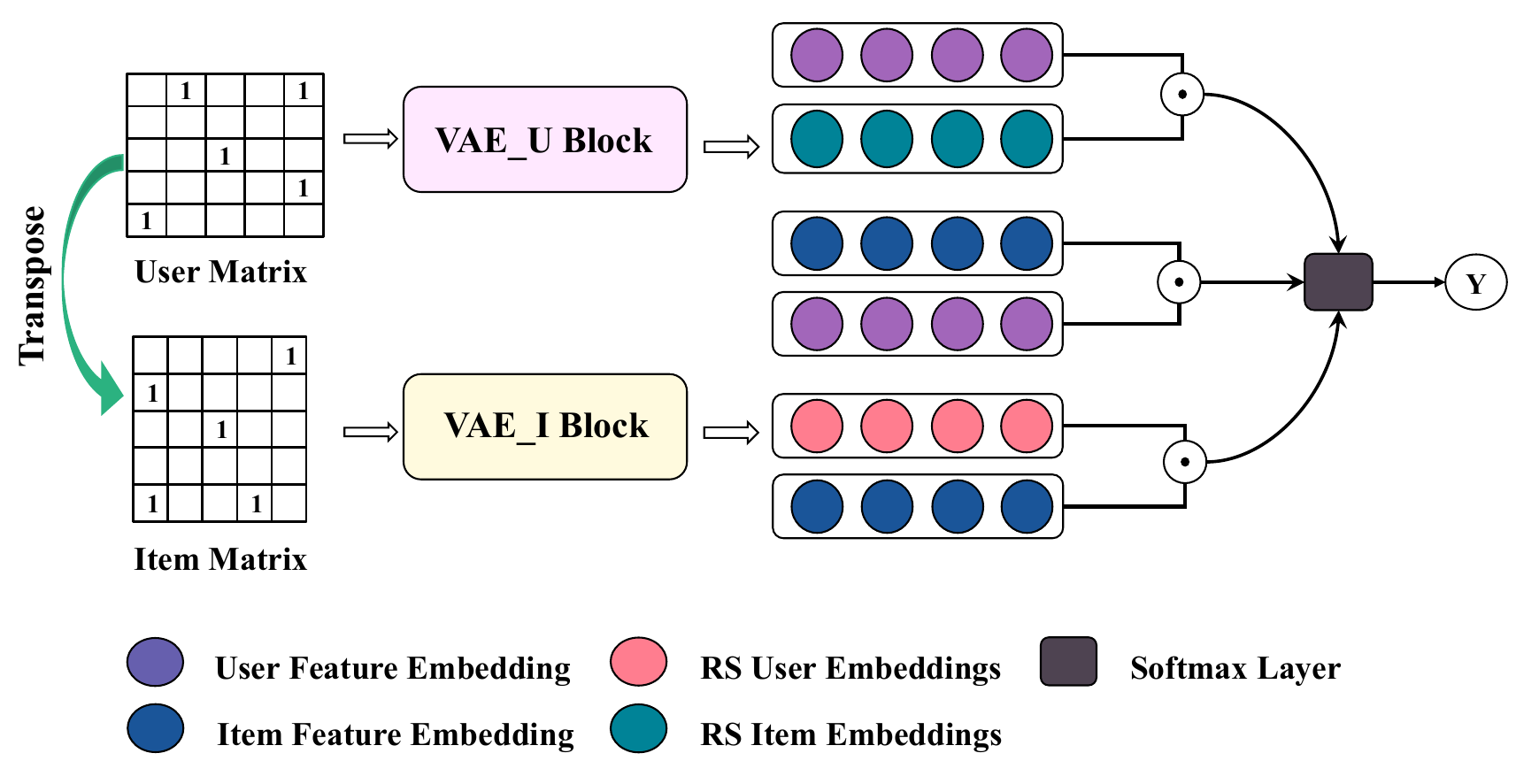}
\end{figure}

\noindent\textbf{True Preference Modeling.} Given the representation of confounders, we can calculate the feedback in the presence of confounders, and obtain the counterfactual preference when there are no confounders as the true preference we want. To make the following analysis more understandable, we use the method of calculating the relevance score of user-item pairs in Matrix Factorization (\textbf{MF}) and treat MF as our target recommendation model. For a user-item pair, we can calculate the relevance score as follows:
\begin{equation}
Score_{u, i} = \mathit{w} \odot \mathit{v} .
\end{equation}
where $w$ and $v$ are the preference of the model capture, embeddings of users, and items respectively. Here are various loss functions for the recommendation task, e.g. BCE and MSE, we choose BCE as the loss function for implicit feedback and MSE for explicit feedback. Then we have the general loss function in the absence of confounders: 
\begin{equation}
    \mathcal{L}_{normal} = \mathbf{BCE}\left(\sigma(Score),Y\right),
\end{equation}
where $\sigma(\centerdot)$ is the Sigmoid function. Next, we introduce unmeasured confounders. As we discussed above, the influence of confounders is divided into two parts corresponding to the user and the item, and similarly, the preference deviations also come from both the user and the item parts, which we need to calculate separately. The formula is shown as follows:
\begin{equation}
Score_{bias}^{U = u} = \textbf{r}_{u} \odot \textbf{v},\quad
Score_{bias}^{I = i} = \textbf{r}_{i} \odot \textbf{w}.
\end{equation}
It is straightforward to explain that the confounders influence the final observed feedback by influencing the performances of users on the item, and the same holds true for the item side. For example, once the recommendation policy decides to expose an item to all users, a higher expected score will be observed due to the higher $Score_{bias}^{I = i}$. The observed feedback is generated by a superposition of the true preferences of the users and confounders mixture, so we need to consider both parts in concert when fitting the observed feedback, which can be formulated as:
\begin{equation}
    \label{eq:10}
    \small
    Score_{bias} =
    \begin{cases}
        Score + Score_{bias}^{I = i} + Score_{bias}^{U = u} & explicit\enspace feedback \\
        Score * Score_{bias}^{I = i} * Score_{bias}^{U = u} & implicit\enspace feedback\\
    \end{cases}.
\end{equation}
When we separate the true preferences of the users from the mixed preferences, we obtain user preferences that are ideally the true preferences of the users, hence the score in Eq. \ref{eq:10} is the counterfactual score in the absence of confounders. Same as $\mathcal{L}_1$, we also use BCE as the loss function for implicit feedback and MSE for explicit feedback, then we have the general loss function in the presence of confounders:
\begin{equation}
    \mathcal{L}_{bias} = \mathbf{BCE}(\sigma(Score_{bias}),Y).
\end{equation}
The Loss function $\mathcal{L}_2$ can help to free the model from the effects of confounders, ensure the modeling of the true preferences of users and items, and hence enhance the performance of the model. In summary, we propose the final loss function of SLFP, which can be formulated as:
\begin{equation}
    \mathcal{L}_{SLFR} = \mathcal{L}_{normal} + \gamma\mathcal{L}_{bias},
\end{equation}
where $\gamma$ is a temperature hyperparameter, used to control the debiasing strength of the model, higher values of $\gamma$ imply stronger debiasing, we will discuss the effect of the value of $\gamma$ in the following experiments. The Framework of SLFR is shown in Figure \ref{fig:4}.
\section{Experiments}
% We conduct experiments in five real-world datasets to evaluate the performance of STPR and other baselines, in comparison with the same benchmark methods. Our experiments aim to answer the following questions:\par
% $\blacklozenge$ Do STPR outperform other debias methods?\par
% $\blacklozenge$ Can STRP address various specific biases?\par
% $\blacklozenge$ Can STPR capture real user preferences compared to other methods?\par
% $\blacklozenge$ What is the impact of the learned representation of latent confounders on the modeling of user preferences?\par
% $\blacklozenge$ Does STPR applicable to other complex recommender system models to improve performance?\par
\begin{table}[!t]
\renewcommand{\arraystretch}{1.2}
\centering
\caption{The statistic of Coat, Yahoo!R3, Ml-1M, KuaiRec, and Reasoner}
\begin{tabular}{@{}lrrrr@{}}
\toprule
             & User & Item & Interactions & parse \\ \midrule
Coat         & 290    & 300    & 6,960           & 92.0\%   \\
Yahoo!R3     & 15,400  & 1,000   & 311,704         & 97.9\%   \\
Movielens 1M & 6,040   & 3,900   & 100,209         & 99.6\%   \\
KuaiRec      & 7,176   & 10,729  & 12,530,806       & 83.7\%   \\
Reasoner     & 1,000   & 4,602   & 19,901          & 99.6\%   \\ \bottomrule
\end{tabular}
\label{tab:1}
\end{table}
\subsection{Experimental Setup}
\noindent\textbf{Datasets.} Our experiments are conducted on five real-world datasets from  Movielens 1M, Yahoo! R3, Coat, KuaiRec\cite{kuairec} and Reasoner. Coat, Yahoo R3, and KuaiRec both have fully observed data as the test set that has the ground truth relevance information, and Movielens 1M as a representative of the common datasets. Following prior works, ratings in Movielens 1M, Yahoo! R3, and Coat are binarized by setting ratings $\geq$ 4 to 1 and the rest to 0. For KuaiRec, We define the interest of a user-item interaction as the watching ratio, which is the ratio of watching time length to the total length, and also binarize it by setting the ratio under 2 to 0, and the rest to 1. Meanwhile, we apply leave-one-out to split training, validation, and test data for all datasets. For Reasoner, we set ratings $\geq$ 4 to 1 and the rest to 0 as the normal train-test set, and use the true user preference label like-unlike as the test set for evaluating the performance of models in capturing real user preference. The properties of datasets are summarized in Table \ref{tab:1}.\\
\noindent\textbf{Baselines.}\textbf{IPS} \cite{ips} adds the standard Inverse Propensity Weight to reweight samples to alleviate item popularity bias. \textbf{CauseE} \cite{cause} applies regularization to force the two sets of embeddings created for unbiased and biased data to be similar. \textbf{DCF} \cite{dcf} uses PMF as a pre-trained model for generating prior distributions. \textbf{DICE} \cite{dice} samples cause-specific data to learn two decomposable embeddings for two causes, user interest, and conformity, to alleviate the popularity bias problem. \textbf{InvPref} \cite{inpref} iteratively decomposes the invariant preference and variant preference from biased observational user behaviors. \textbf{ExposeureMF} \cite{exposure} is a probabilistic model that separately estimates the user preferences and the exposure. \textbf{MACR} \cite{macr} perform counterfactual inference to remove the effect of item popularity.\\
\noindent\textbf{Implementation details.} We implement SLFR and baselines in PyTorch. We take the embedding dimension as 64, and all models are trained with the Adam optimizer via early stopping. We set the learning rate to 1e-3 and the $l_2$-regularization weight to 1e-6. To detect significant differences in SLFR and the best baseline on each data set, we repeat their experiments five times by varying the random seeds, we choose the average performance to report. All ranking metrics are computed at a cutoff K = [10,20,30] for the Top-k recommendation.\par
\subsection{Overall Comparison}
In this section, we compare the performance of the methods in the general debiasing setting. We conduct experiments on Yahoo, Coat, Reasoner, Kuairec, and Ml-1M datasets. The results are shown in Table \ref{tab:2}, we can observe that: Compared to other baselines, the proposed SLFR achieves the best performance with a remarkable improvement for all the metrics across all datasets. Showing the gain in experimental performance due to the separation of user preferences and unmeasured confounders in the latent parameter space demonstrates the superiority of our model. Compared to MF, the methods (e.g. IPS, CauseE) provide extra information to facilitate the model modeling by inferring the prior. Although they can achieve better performance than MF on some datasets (e.g. Kuairec), the high variance problem due to the prediction accuracy of the prior makes the performance of such methods unstable to poor performance than MF on some datasets (e.g. Coat). To overcome the high-variance problem of the inferred prior, DCF uses PMF as a pre-trained model for generating prior distributions, and although this solves the unstable problem in part, its inability to distinguish between confounders and user preferences in the prior leads to the result that the prior it generates remains a mixture, which in turn makes the performance not much different from that of MF across all datasets. Moreover, methods that attempt to disentangle user preferences at the modeling stage achieve stable improvements on all datasets, with InvPref specifically designed for MF being prominent. However, since it leaves the disentanglement of user preferences to the model itself and does not consider the influence of the unmeasured confounders on user preferences, the performance of the model is still sub-optimal. SLFR not only considers the influence of confounders on user preferences but also disentangles them in the latent parameter space to obtain a representation of the confounders for guiding the model, thus guaranteeing further model performance and stability with little increase in model computation.
\begin{table*}[t]
\centering
\caption{Overall performance on five real-world datasets with K = 10 (\%). The best results are highlighted in bold, and the best baseline result in each line is underlined. Higher Recall and NDCG mean better performance.}
\renewcommand{\arraystretch}{0.8}
\begin{tabular}{c|cc|cc|cc|cc|cc}
\toprule
           & \multicolumn{2}{c|}{Coat}       & \multicolumn{2}{c|}{Yahoo}      & \multicolumn{2}{c|}{Reasoner}   & \multicolumn{2}{c|}{Kuairec}     & \multicolumn{2}{c}{Ml-1M}       \\ \midrule
Metric     & Recall         & NDCG           & Recall         & NDCG           & Recall         & NDCG           & Recall          & NDCG           & Recall         & NDCG           \\ \midrule
MF         & 6.563          & 1.755          & 3.534          & 0.660          & 2.961          & 1.036          & 1.848           & 0.359          & 7.719          & 1.332          \\
IPS        & 5.946          & 1.563          & 3.173          & 0.579          & 2.941          & 1.121          & 1.856           & 0.359          & 7.689          & 1.302          \\
CauseE     & 5.136          & 1.622          & 2.429          & 0.437          & 2.826          & 0.655          & 1.852           & 0.360          & 7.383          & 1.302          \\
DCF        & 6.565          & 1.758          & 3.535          & 0.660          & 2.961          & 1.036          & 1.856           & 0.359          & 7.719          & 1.336          \\
DICE       & \uline{8.102}  & \uline{1.795}  & \uline{4.594}  & 0.832          & 3.396          & 0.992          & \uline{1.863}  & \uline{0.369} & 7.949          & \uline{1.374} \\
%ExposureMF & \uline{8.323} & 2.126          & \uline{4.823} & \uline{0.894} & 3.471          & 1.258          & 1.795           & 0.353          & \uline{7.978} & 1.361          \\
%MACR       & 8.071          & 2.15           & 4.817          & 0.883          & 2.043          & 0.763          & 1.856           & 0.359          & 7.921          & 1.369          \\
InvPref    & 4.657          & 1.291          & 4.472          & \uline{0.834}          & \uline{3.584} & \uline{1.279} & 1.847           & 0.357          & 7.918          & 1.369          \\ \midrule
STPR       & \textbf{9.213} & \textbf{2.245} & \textbf{5.470}  & \textbf{0.983} & \textbf{4.196} & \textbf{1.466} & \textbf{1.870} & \textbf{0.373} & \textbf{8.108} & \textbf{1.402} \\ \bottomrule
\end{tabular}
\label{tab:2}
\end{table*}
\begin{table}[tp]
\centering
\caption{Performance of Dealing with Popularity Bias and Exposure Bias on Moviewlens-1M dataset.}
\renewcommand{\arraystretch}{0.8}
\resizebox*{0.8 \linewidth}{!}{
\begin{tabular}{@{}c|ccc|ccc@{}}
\toprule
\multirow{2}{*}{Top K} & \multicolumn{3}{c|}{Recall}                         & \multicolumn{3}{c}{NDCG}                            \\ \cmidrule(l){2-7} 
                       & 10              & 20              & 30              & 10              & 20              & 30              \\ \midrule
MF                     & 0.0772          & 0.1236          & 0.1583          & 0.1332          & 0.1141          & 0.1021          \\
ExposureMF             & \uline{ 0.0807}    & 0.1236          & 0.1642          & 0.1329          & 0.1132          & 0.1007          \\
MACR                   & 0.0476          & 0.0861          & 0.1205          & 0.0884          & 0.0827          & 0.0784          \\
InvPref                & 0.0802          & \uline{ 0.1297}    & \uline{ 0.1674}    & \uline{ 0.1368}    & \uline{ 0.1180}    & \uline{ 0.1061}    \\ \midrule
STPR                   & \textbf{0.0811} & \textbf{0.1338} & \textbf{0.1712} & \textbf{0.1401} & \textbf{0.1212} & \textbf{0.1086} \\ \bottomrule
\end{tabular}}
\label{tab:3}
\end{table}
\subsection{Performance on Debiasing} 
To verify the performance of SLFR on specific biases, we select the two most common biases, popularity bias, and exposure bias. We compare each type of bias with a typical method designed for this bias (e.g. MCAR), while adding a debiasing method designed for MF, InvPref, as a stronger baseline. From the results shown in Table \ref{tab:3}, Methods that focus on specific biases can improve performance, but removing a particular bias does not guarantee that the model will perform as expected, since biases in user preferences result from multiple biases mixed. InvPref achieves generalized debiasing by disentangling user preferences, considering the causes of bias in a comprehensive manner, and not focusing on specific biases, thus achieving better performance than traditional debiasing methods. SLFR further analyzed the deviation of user preferences by separating the confounders and user preferences from a more fundamental latent parameter space to achieve general debiasing, thus showing the best performance.
\begin{table*}[t]
\caption{Overall performance on Real user preference label in Reasoner dataset with K = 10, 20. The best results are highlighted in bold.}
\centering
\renewcommand{\arraystretch}{0.8}
\resizebox*{0.95\linewidth}{!}{
\begin{tabular}{@{}c|c|c|ccccccccc@{}}
\toprule
                          &                         & Top\_K & MF     & IPS    & CauseE & DCF    & DICE   & ExposureMF      & MACR            & InvPref         & STPR            \\ \midrule
\multirow{4}{*}{Reasoner} & \multirow{2}{*}{Recall} & 10     & 0.0714 & 0.1714 & \uline{0.2980} & 0.0714 & 0.0714 & 0.2885          & 0.2649          & {0.2839} & \textbf{0.3232} \\
                          &                         & 20     & 0.3303 & 0.3303 & 0.2980 & 0.3303 & 0.3170 & 0.4035          & 0.5352          & \uline{0.5767} & \textbf{0.6580}  \\ \cmidrule(l){2-12} 
                          & \multirow{2}{*}{NDCG}   & 10     & 0.0110 & 0.0173 & 0.0373 & 0.0110 & 0.0139 & 0.0616          & 0.0828          & \uline{0.1003} & \textbf{0.1518} \\
                          &                         & 20     & 0.2185 & 0.2196 & 0.2407 & 0.2185 & 0.2632 & 0.2991          & \uline{0.4118} & 0.3874          & \textbf{0.4244} \\ \bottomrule
\end{tabular}}
\label{tab:4}
\end{table*}

\begin{figure*}[tp]
\centering
\caption{The impact of the learned representation of latent confounders on the modeling of user preferences}
\includegraphics[width = 0.95\linewidth]{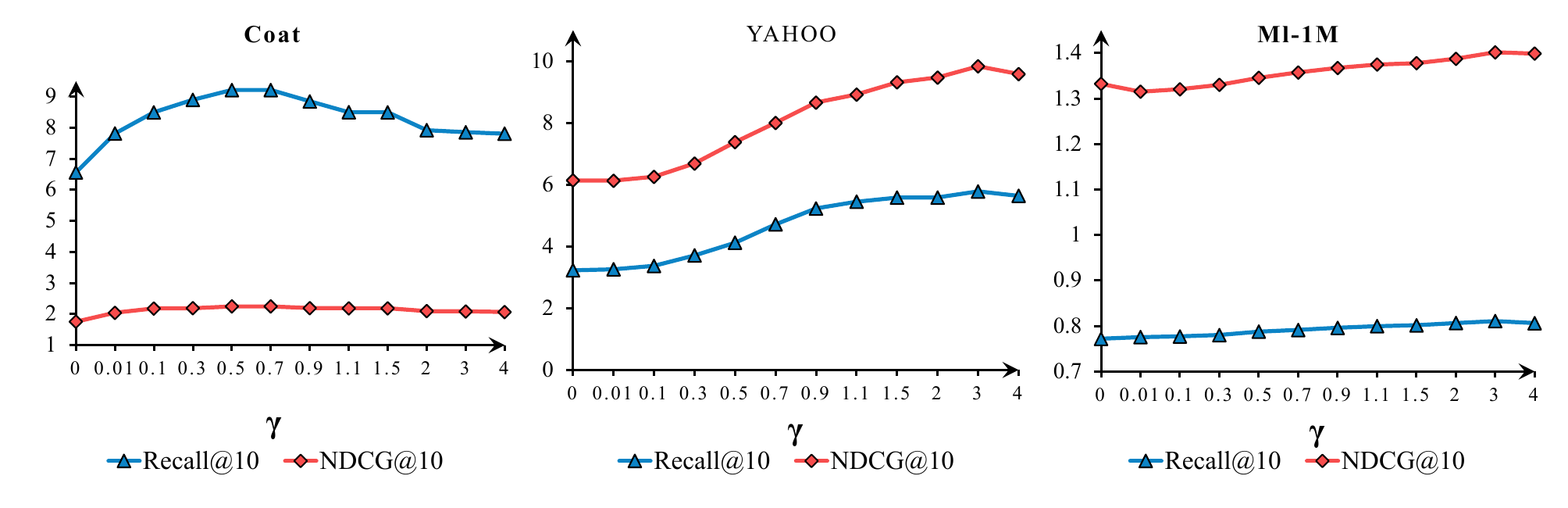}
\label{fig:5}
\end{figure*}
\begin{table}[t]
\centering
\caption{Performance of SLFR on LightGCN, DIN, and SASRec with ML-1M dataset.}
\renewcommand{\arraystretch}{0.8}
\begin{tabular}{@{}c|cc|cc|cc@{}}
\toprule
\multirow{2}{*}{Model} & \multicolumn{2}{c|}{LightGCN} & \multicolumn{2}{c|}{DIN} & \multicolumn{2}{c}{SASRec} \\ \cmidrule(l){2-7} 
                       & -             & SLFR          & -           & SLFR       & -            & SLFR        \\ \midrule
HR@10                  & 0.1562        & 0.1654        & 0.1874      & 0.1933     & 0.2156       & 0.2195      \\
NDCG@10                & 0.1033        & 0.1052        & 0.1095      & 0.1123     & 0.1162       & 0.1183      \\ \bottomrule
\end{tabular}
\label{tab:5}
\end{table}
\subsection{Case Study}
Thanks to the availability of the Reasoner dataset, we are able to have both general feedback data and real user preference labels. We evaluate the performance of all baselines and SLFR on real user preference label data, trained with general feedback data. The results are shown in Table \ref{tab:4}. The methods that fit the data purely perform worse on real labels than on regular data, because the resulting preferences are mixtures with biases that differ from the true preferences or are even diametrically opposite. Even if user preferences are disentangled in the preference modeling phase (e.g. DICE), performance is still not guaranteed. Other debiasing methods can provide some performance guarantees on real label data due to their ability to mitigate the effects of one or more biases, among which InvPref, which uses general debiasing, has outstanding performance. SLFR achieves the best results on the real label dataset, which demonstrates that SLFR performs excellently in capturing the real preferences of users compared to other debiasing methods.
\subsection{Effect of the learned representation of latent confounders}
In this section, we study the representation of unmeasured confounders in our proposed framework. The results are shown in Figure \ref{fig:5}. The performance of SLFR becomes better as the number of $\gamma$ increases, which demonstrates that the representation of unmeasured confounders does help the model to capture the true user preferences. However, the excessive weight gama will lead to the degradation of the model performance, which is because SLFR can only separate the confounding factors and user preferences partly, but cannot reach the ideal state of complete separation of the two, thus the excessive weight will lead to to the model lose part of the real preferences of users and lead to the degradation of performance. We observe that the gamma values are different when the model achieves optimal performance on different datasets, indicating that the deviation of user preferences varies across datasets. The general confounders are not different for different datasets and are caused by the presence of the former recommender system we mentioned above. In other words, the more iterations the current recommender system undergoes when acting as a confounder to influence the user preferences, the larger the deviation from the true user preferences, and hence the smaller the optimal gamma value on the Coat dataset compared to Ml-1M.\par
\subsection{Performance on complex models}
We employ SLFR on LightGCN, DIN, and SASRec to verify the applicability of SLFR to state-of-the-art models, and the results are presented in Table 5. After deploying SLFR, LightGCN, DIN, and SASRec all achieved performance improvements, thus demonstrating that SLFR can still be effective on complex models.
\section{Related Work}
\textbf{Disentanglement}. Disentanglement is commonly described as the problem of extracting the factors of variation responsible for data generation, which are usually considered independent variables associated with a semantic meaning. There has been a lot of outstanding work in the vision domain \cite{betavae}. However, there is very limited work related to the recommender system domain and the work focuses on decoupling user interests, etc., such as varying interests and invariant interests \cite{dice,inpref}. Though they achieve improvements in recommendation performance, there is no guarantee that the recommendations are unbiased, and the learned user preferences are still a mixture of true preferences and other confounders. Different from the previous work, we attempt to separate the influence of confounders by factorizing the latent space in the user preference prediction stage, and thus obtain more accurate user preferences.\par
\noindent\textbf{Causal Recommendation}. Data-driven recommender systems have been widely used in past studies to alleviate the information explosion on the Web. Despite the great success that has been achieved, this data-driven approach still suffers from bias, unfairness \cite{ref4}, and filter bubbles \cite{ref5}. Confounding bias is prevalent in recommender systems due to various confounding factors. For example, item popularity can generate popularity bias and is considered a confounding factor. As causal inference has become a popular approach to eliminate bias in recommender systems and examine the relationships between variables \cite{ref6}, researchers have focused more on the challenge of confounding bias. Compared with previous work, we utilize a proxy variable \cite{ref12} method to obtain information for measuring confounders by factorizing the latent space, based on a generous assumption that satisfies the requirement for a proxy variable that theoretically guarantees the identification of potential outcomes.
\section{Conclusion and Future Work}
In this work, we use causal analysis to thoroughly analyze the causes of user preference deviation and find that the recommender system inherits the influence from confounders and continuously exacerbates the user preference deviation with the recommender system update iterations. To mitigate the influence of unmeasured confounders and former recommender systems on user preferences, we use the former recommender system as their proxy and use variational inference to separate the confounders and user preferences in the latent parameter space to obtain the representation of the confounders. The effectiveness of SLFR demonstrates that separating the confounders can bring performance improvements and more accurate user preferences. However, SLFR cannot explicitly identify certain confounders and their effects, we consider further separation of confounders by further splitting the latent space in future work.
\section{Acknowledge}
This work is supported by the Natural Science Foundation of China for Young Scholars No. 62002132, Jilin Education Science Foundation JJKH20221010KJ, Jilin Science and Technology Research Project 20230101067JC.
\bibliographystyle{splncs04}
\bibliography{reference}
\end{document}